\documentclass[natbib]{svjour3}             % onecolumn (standard format)
\smartqed  % flush right qed marks, e.g. at end of proof
\usepackage{graphicx}
\usepackage{amsmath,amssymb}
\usepackage{lineno}
\usepackage{aas_macros}
\newcommand{\mearth}{M$_{\oplus}$}
\newcommand{\rearth}{R$_{\oplus}$}

\begin{document}

\title{Below One Earth:}
    
\subtitle{The Detection, Formation, and Properties of Subterrestrial Worlds}
    
\titlerunning{Below One Earth Mass}        % if too long for running head
    
\author{E. Sinukoff \and B. Fulton \and L. Scuderi \and E. Gaidos}

\authorrunning{Sinukoff et al.} % if too long for running head
    
\institute{E. Sinukoff \and B. J. Fulton \and L. Scuderi \at
  Institute for Astronomy\\
  University of Hawai`i at M\={a}noa, Honolulu, HI 96822\\  
  \and
  E. Gaidos \at
  Department of Geology \& Geophysics\\
  University of Hawai`i at M\={a}noa, Honolulu, HI 96822\\
  \email{gaidos@hawaii.edu}
}

\date{Received: date / Accepted: date}

\maketitle
\begin{abstract}
  The Solar System includes two planets --- Mercury and Mars ---
  significantly less massive than Earth, and all evidence indicates
  that planets of similar size orbit many stars.  In fact, one of the
  first exoplanets to be discovered is a lunar-mass planet around a
  millisecond pulsar.  Novel classes of exoplanets have inspired new
  ideas about planet formation and evolution, and these ``sub-Earths" should
  be no exception: they include planets with masses between Mars and
  Venus for which there are no Solar System analogs.  Advances in
  astronomical instrumentation and recent space missions have opened
  the sub-Earth frontier for exploration: the \emph{Kepler} mission
  has discovered dozens of confirmed or candidate sub-Earths
  transiting their host stars.  It can detect Mars-size planets around
  its smallest stellar targets, as well as exomoons of comparable
  size.  Although the application of the Doppler method is
  currently limited by instrument stability, future spectrographs may
  detect equivalent planets orbiting close to nearby bright stars.
  Future space-based microlensing missions should be able to
  probe the sub-Earth population on much wider orbits.  A census of
  sub-Earths will complete the reconnaissance of the exoplanet mass
  spectrum and test predictions of planet formation models, including
  whether low-mass M dwarf stars preferentially host the smallest
  planets.  The properties of sub-Earths may reflect their low
  gravity, diverse origins, and environment, but they will be elusive:
  Observations of eclipsing systems by the \emph{James Webb} Space
  Telescope may give us our first clues to the properties of these
  small worlds.
\end{abstract}
    
\section{Uncovering the Sub-Earth Realm}
\label{sec:intro}
    
The detection of Earth-size planets around other stars has long been a
goal of astronomy. The \emph{Kepler} space mission has discovered many
such candidates \citep{Borucki11,Batalha12}, some of which are
confirmed, and seeks to determine the fraction of solar-type stars
that harbor Earth-size planets in their habitable zones
\citep{Borucki10}. While it is human nature to search for analogs of
our home planet, the distribution of objects in our Solar System
extends well below one Earth mass, i.e. Mars, Mercury, the ``dwarf
planets" Pluto and Ceres, and large planetary satellites such as Titan
and Ganymede.  However, the sub-Earth realm remains largely unexplored
around other stars due to the limits of present detection
methods. Like Mars, Europa and Titan, sub-Earth-mass objects in other
planetary systems may prove to be of astrobiological interest. 

For the purposes of this paper, we define sub-Earths or subterrestrial
exoplanets (hereafter STEPs) as planets with radius
$R_{p}<0.95$~\rearth{}.  For a rocky planet this corresponds to a mass
$M_{p}<0.82$~\mearth{}.  Under this definition, a Venus twin
(0.95~\rearth{}, 0.82~\mearth{}) is not a STEP, while analogs to
Mercury (0.38~\rearth{}, 0.055~\mearth{}) and Mars (0.53~\rearth{},
0.11~\mearth{}) are\footnote{We discuss large satellites of
  exoplanets, or ``exomoons'' in \S \ref{subsec:exomoons}.}.
Throughout this paper, we will refer to mass and radius
interchangeably. Conceivably, planets might exist that have a radius
smaller than 0.95 but a mass exceeding that of Earth (e.g. an ``iron
planet''), but we assume a single mass-radius relation appropriate for
an Earth-like composition \citep[i.e.,][see \S
\ref{sec:properties} for more on expected STEP properties]{Valencia07}.
    
\begin{figure}[h]
  \centering
  \includegraphics[scale=0.45]{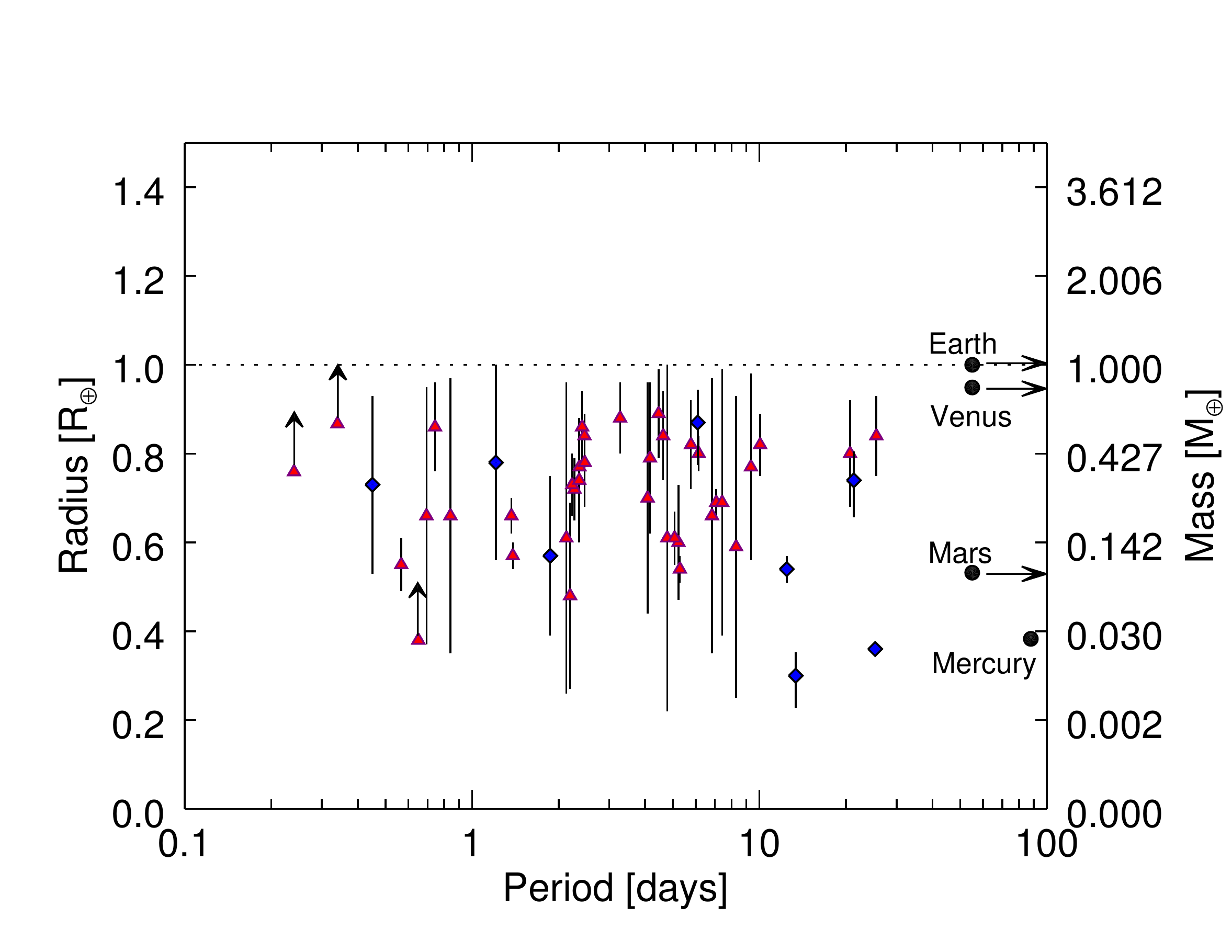}
  \caption{Radii and orbital periods of confirmed (blue diamonds) and
    candidate STEPs (red triangles).  The latter, all but one of which
    are \emph{Kepler} detections, are included only if they have
    estimated radii at least 1$\sigma_{R_p}$ below 1~\rearth{}
    according to the NASA Exoplanet Archive. The void near 1~\rearth{}
    is a result of this criterion.  Where uncertainties are
    unavailable, we include $R < 1$~\rearth{}.  We equate radii and
    mass using a relation for low-mass rocky planets with negligible
    water content \citep{Valencia07}.}
  \label{fig:Rp_vs_P}
\end{figure}

One of the first exoplanets discovered was the 0.02~\mearth{} pulsar
planet PSR~B1257+12A \citep{Wolszczan94}, however the {\it
  Kepler} mission is the primary source of STEP discoveries: As of
April 2013, \emph{Kepler} has discovered 7 STEPs and 36
candidates. Table \ref{tab.planets} lists the parameters of all
reported candidate and confirmed STEPs. Figure \ref{fig:Rp_vs_P} plots
the planets' radii $R_p$ and orbital periods $P$. We include all
confirmed and candidate planets with $R_{p} + \sigma_{R_{p}} <
1$~\rearth{}, where $\sigma_{R_{p}}$ is the uncertainty in radius.
    
STEPs may be very abundant. The distribution of planets rises steeply
with both decreasing mass and decreasing radius down to 3~\rearth{}
\citep{Howard10,Howard12} but appears to be flat from 3~\rearth{} to
1~\rearth{}, the completeness limit of \emph{Kepler}
\citep{Fressin13}.  A data pipeline sensitive to planets as small as
0.5~\rearth{} suggests that the \emph{Kepler} planet size distribution
for $P = $ 5.0--10.8 days either remains flat or increases below
1~\rearth{} \citep{Petigura13}. Although the detection rate in this region of
parameter space is $<$50\%, these findings indicate that STEPs are
relatively common.
        
STEPs occupy diverse environments. Three STEPs orbit the M dwarf
Kepler-42 \citep{Muirhead12a}. Kepler-20e is part of a five-planet
system that includes three gas giants and an Earth-size planet
\citep{Fressin12}. Contrary to the configuration of the Solar System,
the two smallest planets of Kepler-20 orbit amongst the giants. KOIs
55.01 and 55.02 orbit within 0.008~AU of a B subdwarf and have day side
temperatures exceeding 8000~K, allowing \emph{Kepler} to detect their
thermal emission at visible wavelengths \citep{Charpinet11}. They
somehow survived or avoided engulfment by the star during its red
giant phase. An object transiting {\it Kepler} star KIC~12557548 every
15.7~hr is thought to be a disintegrating Mercury-size planet
surrounded by a cloud of dust \citep{Rappaport12}.
    
STEPs are part of the complete picture of planet formation and
evolution. Although planet formation is thought to be a stochastic
process, statistical quantities and occurrence patterns, e.g. mass
distribution or metallicity correlation, presumably reflect underlying
processes common to all systems.  Any planet formation theory is
incomplete if it cannot account for such trends in the sub-Earth
population. Historically, the study of exoplanets in previously
unexplored regions of parameter space has provoked new ideas: The
discovery of ``hot Jupiters'' \citep{Mayor95} led to proposals for
orbital migration \citep{Lin96}, that of ``super-Earths''
\citep{Rivera05} kindled interest in volatile-rich ``ocean planets''
\citep{Kuchner03,Leger04}, and the hot rocky planet CoRoT-7b
\citep{Queloz09} stimulated the concept of ``lava planets''
\citep{Leger11} and ``Chthonian" planets, the remnant cores of
evaporated gas giants \citep{Hebrard04}.  STEPS should likewise expand
our appreciation for --- and demand the explanation of --- the diverse
outcomes of planet formation.
        
In this review, we address the capability of both the \emph{Kepler}
space mission (\S \ref{sec:kepler}) and ground-based Doppler
observations (\S \ref{sec:rv}) to detect STEPs.  In \S \ref{sec:other}
we consider the original (pulsar timing) and one future (microlensing)
method by which STEPs can be detected, as well as the potential for
discovery of exomoons.  In \S \ref{sec:formation} we
discuss the predictions of planet formation theory, and in \S
\ref{sec:properties} we speculate on the properties of STEPs and how
they might be established by follow-up observations.  We summarize our
conclusions and recommend future studies in \S \ref{sec:discussion}.
    
\begin{table}[h]
  \centering
  \begin{minipage}{130mm}
    \caption{Confirmed and Candidate Sub-Earths (STEPs)$^a$} 
    \begin{tabular}{lrrrll}
      \hline\noalign{\smallskip}
      Name & R$_{p}$ (\rearth{}) & M$_{p}$ (\mearth{})\footnote{Assuming M/\mearth{}=(R/\rearth{})$^{3.817}$ appropriate for an Earth-like composition \citep{Valencia07}.} & Period (d) & Method & Reference\footnote{BA13: \citet{Barclay13}, BO13: \citet{Borucki13}, C11: \citet{Charpinet11}, F12: \citet{Fressin12}, M12: \citet{Muirhead12a}, NEA13:  NASA Exoplanet Archive (January 2013), R12: \citet{Rappaport12}, S12: \citet{Stevenson12}, W94: \citet{Wolszczan94}.}\\
      \noalign{\smallskip}\hline\noalign{\smallskip}
      \multicolumn{6}{c}{Confirmed planets}\\
      Kepler-20e \footnote{\emph{Kepler} detections include Quarters 1-8 observations.  We exclude \emph{Kepler} candidates with Multiple Event Statistic (a measure of signal-to-noise) $<$7, which are statistically unreliable detections \citep{Jenkins10}. We omit targets that are unclassified in the \emph{Kepler} Input Catalogue \citep{Brown11}.}
      & 0.87 & 0.59 & 6.10 & transit & F12 \\
      Kepler-37b & 0.30 & 0.01 & 13.37 & transit & BA13 \\
      Kepler-37c & 0.74 & 0.32 & 21.30 & transit & BA13 \\
      Kepler-42b & 0.78 & 0.39 & 1.21 & transit & M12 \\ 
      Kepler-42c & 0.73 & 0.30 & 0.45 & transit & M12 \\ 
      Kepler-42d & 0.57 & 0.12 & 1.87 & transit & M12 \\
      Kepler-62c & 0.54 & 0.10 & 12.44 & transit & BO13 \\
      PSR B1257+12 A & 0.36 & 0.02 & 25.27 & pulsar & W94 \\ 
      \noalign{\smallskip}\hline\noalign{\smallskip}
      \multicolumn{6}{c}{Candidate planets\footnote{Candidate planets have estimated R$_{p} <$ 1.0~\rearth{} \ to within 1$\sigma$.}}\\
      KIC 12557548b\footnote{Quoted $M_p$ and $R_p$ are lower limits.} & 0.38 & 0.03 & 0.65 & transit & R12 \\
      KOI 55.01$^{e}$ & 0.76 & 0.35 & 0.24 & transit & C11 \\ 
      KOI 55.02$^{e}$ & 0.87 & 0.59 & 0.34 & transit & C11 \\ 
      KOI 82.04 & 0.70 & 0.26 & 7.07 & transit & NEA13 \\
      KOI 82.05 & 0.52 & 0.08 & 5.29 & transit & NEA13 \\ 
      KOI 251.02 & 0.82 & 0.47 & 5.77 & transit & NEA13 \\
      KOI 283.02 & 0.84 & 0.51 & 25.52 & transit & NEA13 \\
      KOI 321.02 & 0.84 & 0.51 & 4.62 & transit & NEA13 \\
      KOI 430.02 & 0.77 & 0.37 & 9.34 & transit & NEA13 \\ %
      KOI 568.02 & 0.74 & 0.32 & 2.36 & transit & NEA13 \\
      KOI 605.02 & 0.61 & 0.15 & 5.07 & transit & NEA13 \\
      KOI 672.03 & 0.55 & 0.10 & 0.57 & transit & NEA13 \\
      KOI 952.05 & 0.86 & 0.56 & 0.74 & transit & NEA13 \\ %
      KOI 1499.02 & 0.66 & 0.20 & 0.84 & transit & NEA13 \\ %
      KOI 1612.01 & 0.78 & 0.39 & 2.47 & transit & NEA13 \\
      KOI 1618.01 & 0.77 & 0.37 & 2.36 & transit & NEA13 \\
      KOI 1619.01 & 0.80 & 0.43 & 20.67 & transit & NEA13 \\
      KOI 1692.02 & 0.84 & 0.51 & 2.46 & transit & NEA13 \\
      KOI 1964.01 & 0.73 & 0.30 & 2.23 & transit & NEA13 \\
      KOI 1977.02 & 0.69 & 0.24 & 7.42 & transit & NEA13 \\ 
      KOI 2006.01 & 0.88 & 0.61 & 3.27 & transit & NEA13 \\ %
      KOI 2013.01 & 0.86 & 0.56 & 2.41 & transit & NEA13 \\
      KOI 2029.02 & 0.82 & 0.47 & 10.06 & transit & NEA13 \\
      KOI 2059.01 & 0.80 & 0.43 & 6.15 & transit & NEA13 \\
      KOI 2079.01 & 0.66 & 0.20 & 0.69 & transit & NEA13 \\ %
      KOI 2169.04 & 0.50 & 0.07 & 2.19 & transit & NEA13 \\ 
      KOI 2247.01 & 0.89 & 0.64 & 4.46 & transit & NEA13 \\ %
      KOI 2421.01 & 0.72 & 0.29 & 2.27 & transit & NEA13 \\
      KOI 2426.01 & 0.79 & 0.41 & 4.16 & transit & NEA13 \\ %
      KOI 2527.01 & 0.57 & 0.12 & 1.39 & transit & NEA13 \\
      KOI 2657.01 & 0.60 & 0.14 & 5.22 & transit & NEA13 \\
      KOI 2693.01 & 0.70 & 0.26 & 4.08 & transit & NEA13 \\ %
      KOI 2693.03 & 0.66 & 0.20 & 6.83 & transit & NEA13 \\
      KOI 2792.01 & 0.61 & 0.15 & 2.13 & transit & NEA13 \\ %
      KOI 2838.02 & 0.61 & 0.15 & 4.77 & transit & NEA13 \\ %
      KOI 3083.03 & 0.59 & 0.13 & 8.29 & transit & NEA13 \\
      UCF-1.01 & 0.66 & 0.20 & 1.37 & transit & S12 \\ 
      UCF-1.02 & 0.65 & 0.19 & --- & transit & S12 \\
    \label{tab.planets}
    
  \end{tabular}
\end{minipage}
\end{table} 
    
\section{Detection of Sub-Earths by \emph{Kepler}}
\label{sec:kepler}

Since most of the currently known sub-Earth planet candidates were
discovered by \emph{Kepler}, it is useful to study the sensitivity of
\emph{Kepler} observations to such planets. This will provide an
estimate of the number of stars in the \emph{Kepler} sample around
which the mission could detect transiting sub-Earth planets as well as
identify those stars most suitable for such a search.

The \emph{Kepler} spacecraft was launched in 2009 with the primary
goal of discovering an Earth-size exoplanet in the habitable zone of a
solar-type star \citep{Borucki10,Koch10}.  As of January 2013, the
\emph{Kepler} mission has discovered 105 bona fide planets and more
than 2700 planetary candidates using the transit detection method,
i.e. by detecting the decrease in flux as the planet passes in front
of its host star. Most candidates are likely to be planets
\citep{Colon12, Morton12, Fressin13}, but the stars are either too
faint or the planets too small to be confirmed by the radial velocity
method (\S \ref{sec:rv}). The unsurpassed precision of \emph{Kepler}
photometry and the fact that the transit method is sensitive to the
cross-section ($\propto R_{p}^{2}$) of the planet, not its mass
\citep[$\propto R_{p}^{4}$,][]{Valencia07} makes \emph{Kepler} our
most powerful tool for detecting STEPs.
    
\subsection{Direct transit detection \label{subsec:transit}}
    
We assess the ability of \emph{Kepler} to directly detect STEPs via
transits of their host stars. The transit signal is proportional to
$(R_{p}/R_{*})^{2}$, where $R_*$ is the radius of the star.  At a
given detection limit for a transit signal, smaller planets can be
found around smaller stars.  For example, a 0.5 $R_{\oplus}$ planet
produces a signal of $\sim$20 parts per million (ppm) if it transits a
G5 dwarf, but $\sim$50 ppm if it transits an M2 dwarf.  All else being
equal, late-type M dwarf stars should be more desirable targets for
searches for STEPs.  We first consider the fraction of sub-Earths that
would be detected around the planet-hosting \emph{Kepler} M dwarfs
characterized by \citet{Muirhead12b} (hereafter M12). These stars have
radii and masses inferred from a comparison of
spectrocopically-determined effective temperatures $T_{\rm eff}$ and
metallicities with stellar evolution models.  These allow us to
convert a detection threshold into equivalent planet
radii. Furthermore, these stars are likely to host additional planets
\citep{Wright09,Lissauer11}, and these planets are also likely to
transit because of orbital coplanarity
\citep{Sanchis-Ojeda12,Hirano12}.
    
The radius of the smallest detectable planet $R_p$ with an orbital
period $P$ observed for a time $t_{\rm obs}$ is:
\begin{equation}
  R_p = R_*\sqrt{\left(\frac{P}{t_{\rm obs}}\right)^{1/2} {\rm S/N}\cdot {\rm CDPP}_{d}},
  \label{eqn:SN}
\end{equation}
where the threshhold signal to noise S/N for detection is 7.1
\citep{Jenkins10,Tenenbaum12}, and CDPP is the effective Combined
Differential Photometric Precision over a time interval $d$
\citep{Koch10}.  CDPP$_d$ is a measure of the noise of a light curve
within a specified time interval $d$ and is similar to the standard
deviation of the photometry binned over that interval.
    
\begin{figure}[h]
  \centering
  \includegraphics[scale=0.32]{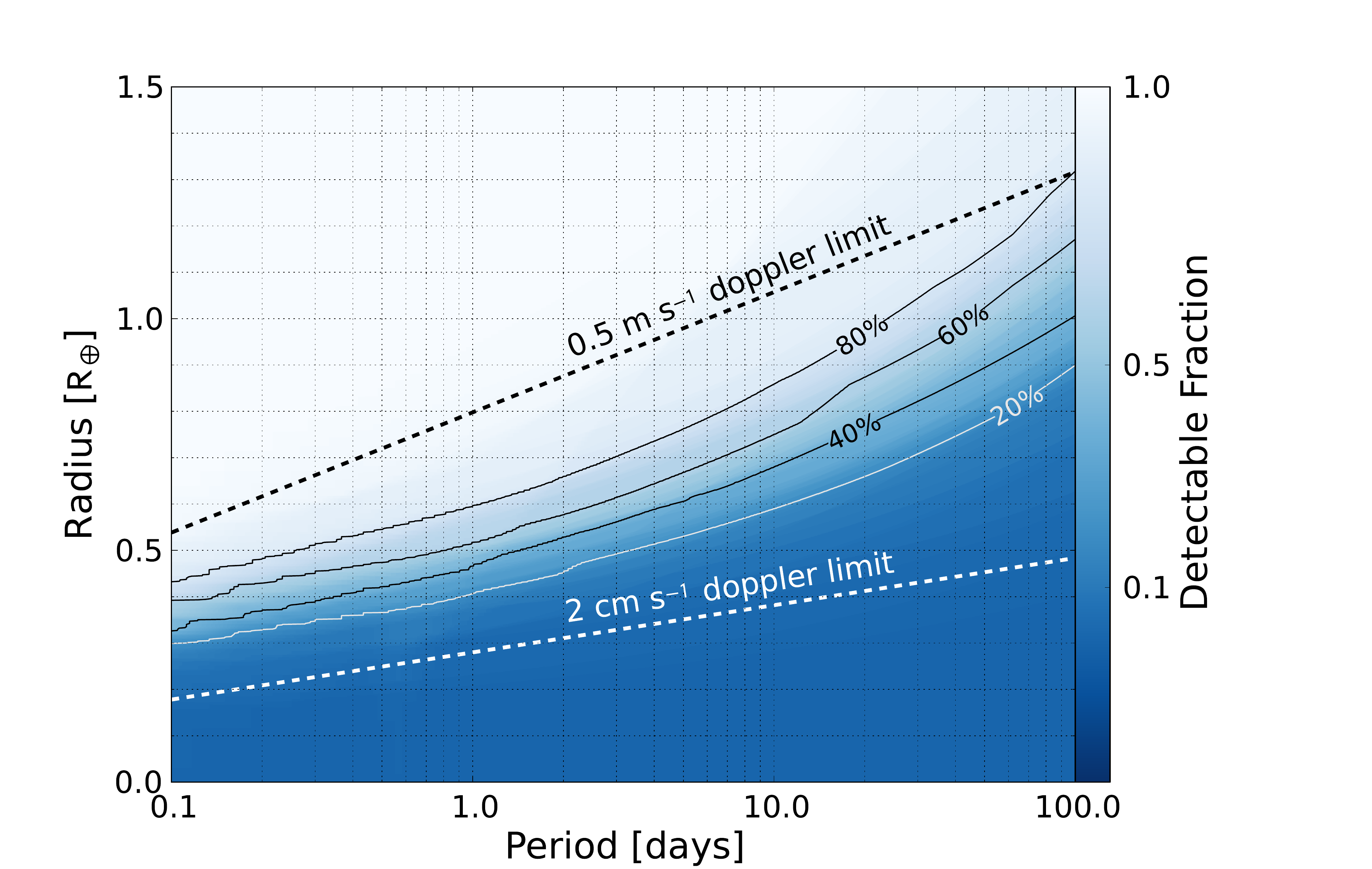}
  \caption{Detectable fraction vs. period and planet radius for
    hypothetical additional planets orbiting planet-hosting stars in
    the M12 catalogue.  All planets are assumed to be on coplanar
    orbits and transiting. The dashed lines show the best current
    radial velocity capability (0.5~m~s$^{-1}$) and future radial
    velocity capability (2~cm~s$^{-1}$) for instruments such as CODEX
    (see \S \ref{sec:rv}) assuming the median stellar mass (0.53
    M$_{\odot}$) of the M12 sample and a planet mass-to-radius
    relation for rocky planets from \citet{Valencia07}. The detectable
    fraction is defined as the fraction of M12 stars around which a
    planet of a given radius and period would produce a detectable
    (7.1 $\sigma$) transit signal over the course of the extended
    \emph{Kepler} mission (6.8~yr, see section 2.1).}
  \label{fig:detectability}
\end{figure}

To determine the sensitivity of the survey to sub-Earth planets around the
M12 stars, we first calculate the transit duration ($d$) for a range
of possible planet orbital periods using the stellar masses and radii
from M12.  To determine CDPP$_d$ we fit a second-order polynomial in
$1/\sqrt{d}$ to the $d=3$, 6, and 12 hour CDPP values of each of the
$\sim$168,000 stars observed by \emph{Kepler}.  Assuming that every
planet transits, and that the observing timespan is equal to the total
length of the \emph{Kepler} primary plus extended mission\footnote{The
  extended mission may be terminated due to the failure of a second
  reaction wheel on the spacecraft.}  ($t_{\rm obs} = 6.8$~yr),
\emph{Kepler} should be able to detect transiting sub-Earth-size
planets with periods as long as $\sim$60 days around $\sim$50\% of the
stars in the M12 catalogue (Fig. \ref{fig:detectability}).
    
Although the transit signal for a given planet size is inversely
proportional to the square of the stellar radius, the transit
signal-to-noise also depends on the noise due to both intrinsic
stellar variability and photometric error \citep{Gilliland11}.  The
smallest planets can be detected around the smallest, brightest, and
most intrinsically quiet stars.  All of the stars in the M12 catalogue
are early- to mid-M-type stars: \citet{Gilliland11} shows that only
7\% of M dwarfs, but 76\% of G5 dwarfs have CDPP$_{6} < 50$~ppm.  In
addition, late-type stars are much less luminous and thus
under-represented in the magnitude-limited \emph{Kepler} survey.
    
The results of \citet{Gilliland11} motivate us to identify the subset
of \emph{Kepler} targets that are best suited for detecting transits
of sub-Earth-sized objects.  These same arguments will also apply to
any future space-based transit survey if the photometric precision is
limited by stellar variability.  Since transit signal-to-noise scales
inversely with the product of $R_{*}^{2}$ and CDPP, we define a
parameter $D \equiv R_{*}^{2}$CDPP$_{6}$ to identify the most suitable
stars for which to search for small planets. We use CDPP$_{6}$ as our
reference because the corresponding orbital period of a transiting
planet is 40~d, within the range considered here.  (The other
available precision metrics are for 3~hr and 12~hr, corresponding to
orbital periods of 5~d and 320~d).

\begin{figure}[h]
  \centering
  \includegraphics[scale=0.3]{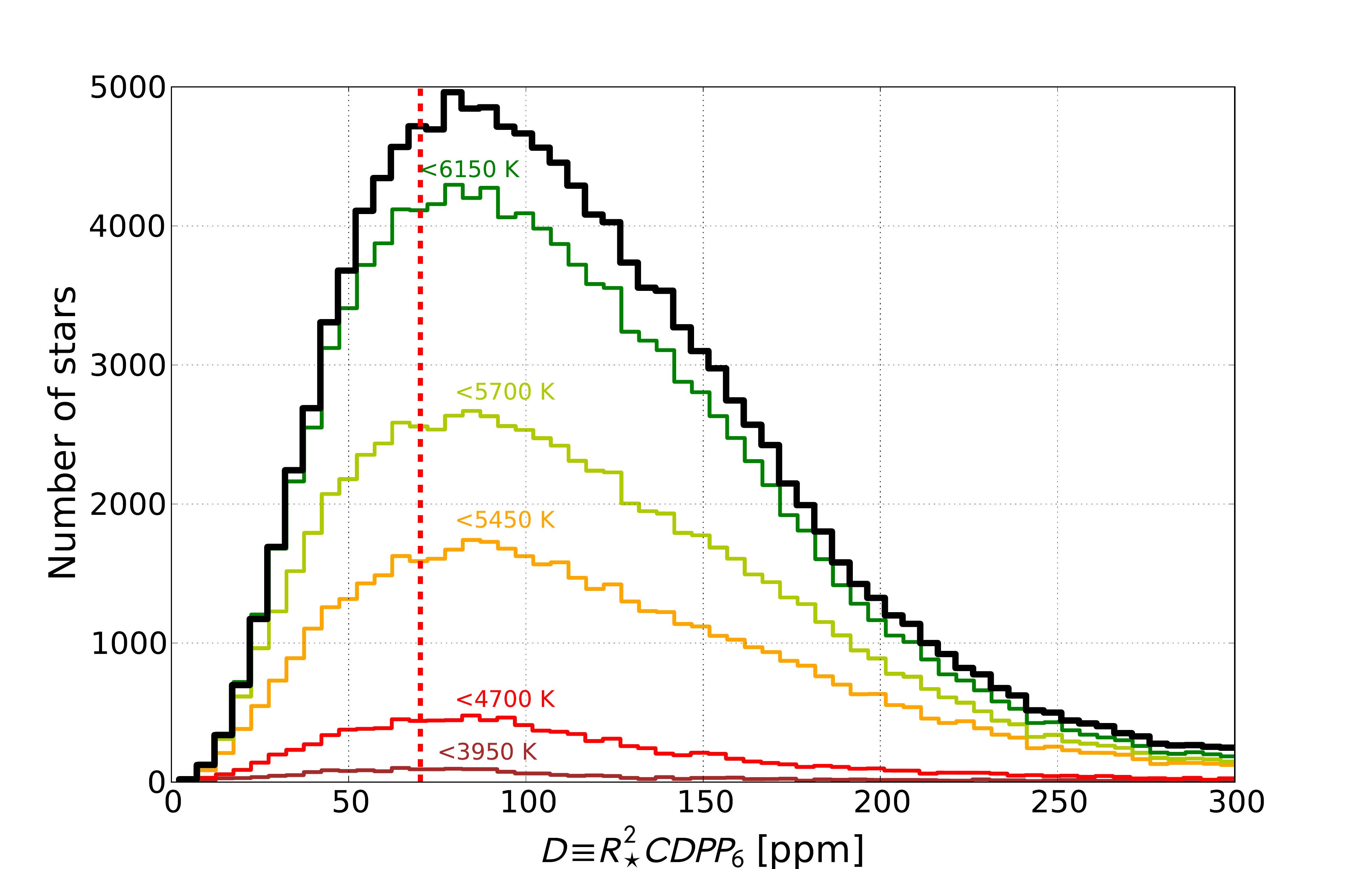}
  \caption{Distribution of $D\equiv R_{*}^{2}CDPP_{6}$ for all
    $\sim$168,000 stars observed by \emph{Kepler} in Quarter 10. The
    colored lines are cumulative bins of stars grouped by $T_{\rm
      eff}$ and the black line is the distribution of the entire
    sample. The vertical dashed red line indicates the 25th percentile
    cut of stars used to produce Fig. \ref{fig:detectprob}.}
  \label{fig:cdppdist}
\end{figure}

Smaller planets can be detected around stars with lower values of $D$.
Figure \ref{fig:cdppdist} shows the distribution of $D$ for those
stars observed by \emph{Kepler}, binned by the $T_{\rm eff}$ reported
in the Kepler Input Catalogue \citep[KIC,][]{Brown11}.  The curves are
cumulative with $T_{\rm eff}$ from coolest to hottest: the uppermost
curve is the total over all $T_{\rm eff}$.  These distributions peak
near $D \approx 80$~ppm, close to the signal from the single transit
of an Earth twin around a solar analog (84~ppm). This means that the
signal-to-noise of individual transits of an Earth twin would be
$\sim$1 and highlights the challenge of detecting such a planet with
\emph{Kepler}.  Planets on shorter-period orbits will produce more
transits and are more readily detected.  About one quarter of all {\it
  Kepler} targets have $D < 70$~ppm.  Figure \ref{fig:cdppdist}
indicates that no particular spectral type is optimal, although stars
with 5450 K $< T_{\rm eff} < 5700$ K have a $D$ distribution slightly
skewed toward lower values.  This contrasts with the common perception
that M dwarfs are favorable targets because, among \emph{Kepler}
stars, M dwarfs are fainter and photometrically noisier. However, the
stellar radii of very low-mass stars in the KIC catalogue are
systematically too large \citep{Muirhead12b,Mann12}.  If the radii
were corrected, this would push the distribution of $D$ for the
M-dwarfs towards smaller values of $D$, making them slightly better
targets.
    
\begin{figure}[h]
  \centering
  \includegraphics[scale=0.32]{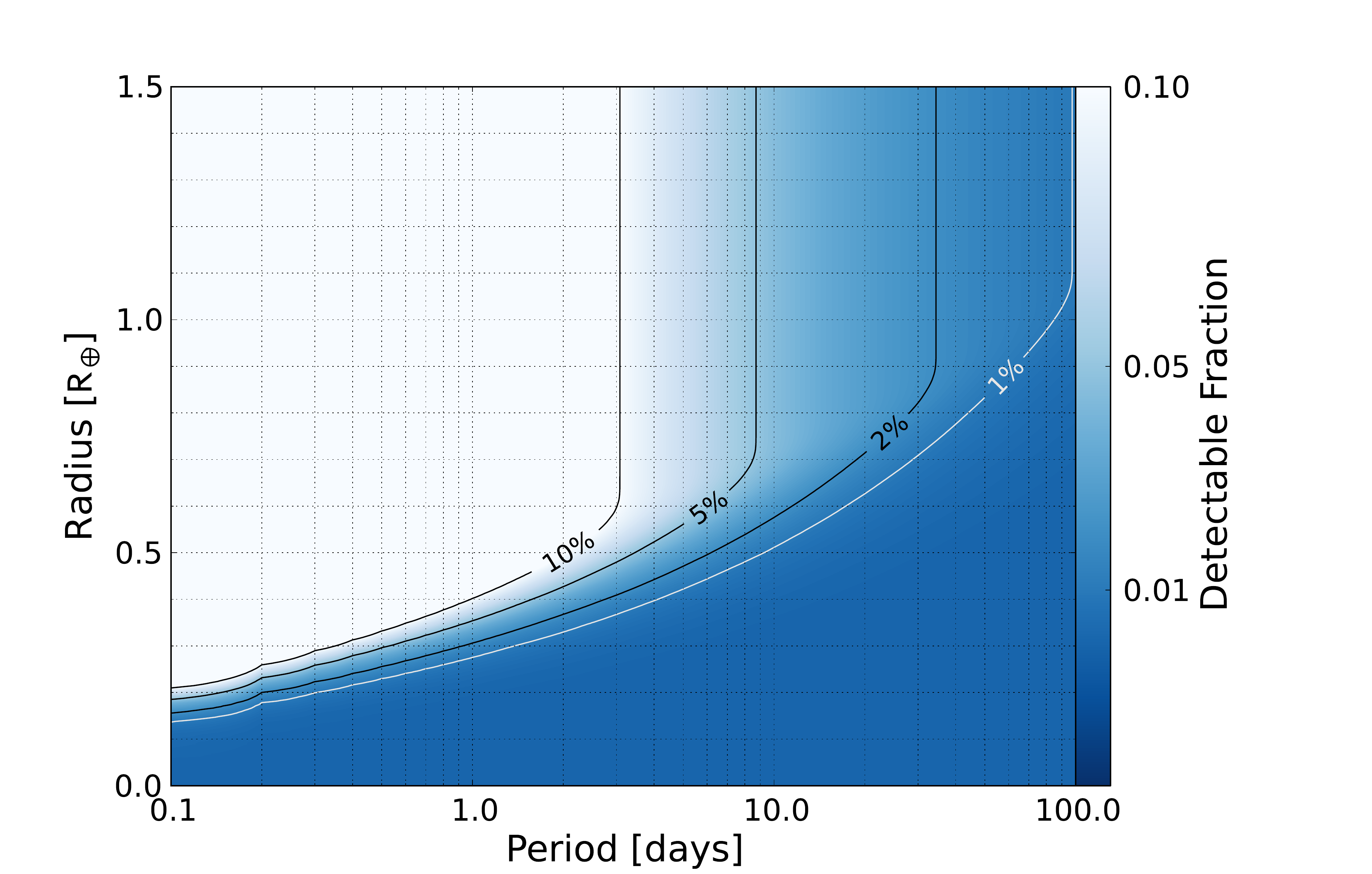}
  \caption{Detectable planet fraction vs. period and planet radius of the
    25\% most detection-favorable \emph{Kepler} targets, i.e. smallest
    $D\equiv R_{*}^{2}CDPP_{6}$, accounting for transit
    probability. The sharp rise of the contours (decrease in
    detectability) at long periods is a result of the geometric
    probability of transit being very low for long-period planets. The
    detectable fraction is defined as the fraction of KIC stars around
    which a planet of a given radius and period would produce a
    detectable (7.1 $\sigma$) transit signal (see Section 2.1).}
  \label{fig:detectprob}
\end{figure}

In order to estimate the fraction of stars around which sub-Earth-size
planets are detectable in the Kepler photometry via transits, we
select the 25\% (32,721) of stars with the smallest values of $D$ and
calculate the percentage of sub-Earth planets that would be detected
if each star was orbited by a planet with a given period.  We assume
isotropically-oriented orbits, such that the transit probability is
$R_*/a$, where $a$ is the semi-major axis, using $R_{*}$ from the KIC
and the scaling relation $M_{*}\sim R_{*}^{0.8}$ \citep{Cox00} to
determine $a$ from $P$.  Figure \ref{fig:detectprob} shows that {\it
  Kepler} should find $\sim1\%$ of planets with $R \sim$ 1~\rearth{}
out to $P \sim 80$~d around these stars.  Thirty-six sub-Earths found
by \emph{Kepler} orbit at $1 < P < 10$~d where the detection
efficiency is $\sim$5-10\% (Fig. \ref{fig:Rp_vs_P}).
    
\subsection{Detection by transit timing variations \label{subsec:ttv}}
    
Additional, non-transiting planets can be detected when mutual
gravitational perturbations cause sufficient variation in the
ephemeris of the transiting planet
\citep{MiraldaEscude02,Holman05,Agol05}.  The amplitude of these
transit timing variations (TTVs) in the case of an inner transiting
planet being perturbed by a longer period companion is
\citep{Holman05}:
\begin{equation}
  \label{eqn:ttv}
  \Delta t \simeq \frac{45\pi}{16}\left(\frac{M_{p}}{M_{*}}\right)\frac{P_{1}\alpha_{e}^{3}}{(1-\sqrt{2}\alpha_{e}^{3/2})^{2}},
\end{equation}
where the subscripts 1 and 2 indicate parameters of the transiting and
perturbing planet, respectively, $P$ is the orbital period, $M_{p}$ is
the perturbing planet mass, $M_{*}$ is the mass of the host star,
$\alpha_{e} = a_{1}/[a_{2}(1-e_{2})]$, and $e$ is the
eccentricity. $\Delta t$ depends on the mass of the perturber (not the
transiting planet), and sub-Earth-mass perturbers will produce only a
very small TTV signal.  The amplitude of a TTV signal also depends on
the period ratio and orbital eccentricities of the two planets and is
maximized when the periods are commensurate.  Dynamical simulations
show that the maximum TTV signal from an Earth-mass planet will be
$\sim$20~s \citep{Holman05}.
    
\emph{Kepler} can measure the time of transit center with a precision
of $\sim$20~s for the deepest transits of the brightest stars, but for
the majority of stars the precision is much worse \citep{Ford11}. The
highest precision that has ever been achieved is $\sim$5~s using the
\emph{Hubble} Space Telescope \citep{Brown01,Pont07}, and ground-based
observations can achieve precisions of $\sim$60~s \citep{Fulton11,
  Maciejewski13}. Although \emph{Hubble} may be able to detect TTVs
due to sub-Earths, the telescope's short observing window (due to its
low Earth orbit) makes it less than ideal for this type of
observation. It seems unlikely that TTVs will be a viable method to
detect STEPs until the advent of a more capable observatory such as
the \emph{James Webb} Space Telescope (JWST).
    
\section{Doppler detection}
\label{sec:rv}
    
The radial velocity (RV) or Doppler method was used to discover and
confirm the first exoplanet around a main-sequence star
\citep{Mayor95}, and is responsible for nearly half of all exoplanet
discoveries to date \citep{Schneider11}. Although this statistic is
changing because of the success of transit surveys such as
\emph{Kepler}, RV measurements are required to rule out certain false
positive scenarios and measure planetary mass.  For a circular orbit, RV
semi-amplitude $K$ scales as
\begin{equation}
  K \approx 64\frac{M_{p}\sin i}{M_{\oplus}} \left(\frac{P}{1 {~\rm d}}\right)^{-1/3} \left(\frac{M_*}{M_{\odot}}\right)^{-2/3} {\rm cm~s}^{-1}.
\end{equation}
where \emph{i} is the orbital inclination of the planet with respect
to the plane of the sky. As of January 2013, the smallest reported
signal is 51 cm s$^{-1}$ from a planet with a minimum mass of
1.3~\mearth{} on a 3.2~d orbit around $\alpha$ Cen B \citep[][but see
\citet{Hatzes13}]{Dumusque12}. This detection was made by the High
Accuracy Radial velocity Planet Searcher (HARPS) instrument installed
on the ESO La Silla 3.6~m telescope in Chile, which represents the
state of the art in operational spectrographs.  Other instruments
achieve an RV stability in the 1--3~m~s$^{-1}$ range (Table
\ref{tb:rvins}).  This performance falls well short of what is needed
to detect Earths or sub-Earths with $P \gg 1$~d around solar-type
stars, but leaves open the possibility of discovering or confirming
``hot" STEPs on extremely close orbits ($P \sim 1$~d) around M dwarfs
(Fig. \ref{fig:detectability}).
    
Detection of STEPs at larger orbital distances will require greatly
improved sensitivity: A 0.5 \mearth{} planet on a 10~d orbit around a
$0.53 M_{\odot}$ star produces a maximum Doppler signal of $\sim22$ cm
s$^{-1}$, or roughly half that of the current best precision reported
by HARPS. The same planet with the same orbit around a solar-type star
would produce a Doppler signal of only $\sim14$ cm s$^{-1}$.  Although
this precision is beyond the abilities of current instruments, there
are already plans in place to improve the performance of existing
instruments, such as HARPS, and to build new instruments which will
achieve the precision needed to detect STEPs. In order to do this, new
instruments must overcome both instrumental and stellar noise. The
solutions come from multiple approaches, and we address each below.
    
Spectrographs mounted directly on telescopes experience flexure,
pressure variations, and temperature variations that produce
systematic errors. These effects can be minimized
by placing the instrument in a temperature-stabilized dewar fed by a
fiber from the telescope, as is done with HARPS \citep{Lovis06}.
Imaging a star directly onto a spectrograph slit engenders noise from
guiding errors and changes in the point spread function (PSF)
\citep{Valenti95,Endl00}. These issues can be partially addressed by
high-cadence pointing corrections, but a more elegant solution is to
stabilize the PSF by transmitting the light to the instrument by a
fiber \citep[e.g.,][]{Spronck12,Bouchy13}.  Isolating and finely
controlling the environment of the instrument is necessary to maintain
both short-term (single observation) and long-term (survey-spanning)
instrumental precision.

Another source of error is the wavelength calibrator against which
Doppler shifts are measured.  A molecular iodine gas absorption cell,
placed in the beamline, provides a forest of fiducial absorption lines
at $\lambda < 650$~nm \citep{Butler96}.  While iodine works well for
observations of solar-type stars, which have significant signal at
blue wavelengths, it becomes a limiting factor for Doppler
observations of M dwarfs, which have peak emission at redder
($\lambda>800$~nm) wavelengths.  An alternative gas is ammonia, which
has a large number of lines in the $K$ (2.2 $\mu$m) band
\citep{Reiners10}.  An ammonia gas cell is used with the CRyogenic
high-resolution InfraRed Echelle Spectrograph (CRIRES) at the ESO Very
Large Telescope (VLT) for radial velocity searches for planets around
M dwarfs \citep{Bean10}.
    
There are also alternatives to gas absorption cells.  The HARPS
spectrograph uses the emission lines from a thorium-argon lamp, but
such lamps also have fewer lines in the near infrared and the light
from the lamp does not follow the exact same path as that from the
star.  A laser frequency ``comb" combined with an etalon
interferometer can create a uniform ladder of equally bright emission
lines across a selectable wavelength range. \citet{Steinmetz08}
suggest that, with more development, laser combs should permit RV
measurements with a precision of $\sim$1 cm s$^{-1}$.
    
The ultimate limit to the Doppler method is intrinsic stellar noise or
``jitter" from granulation, oscillations, plages, and star spots.  One
strategy is to average over these noise terms.  \citet{Dumusque11a}
conclude that a scheme where three 10-minute spectra are obtained 2
hours apart on each of 10 nights per month yields the best radial
velocity precision and minimizes problems from stellar variability.
    
Forthcoming instruments will take advantage of these technologies and
strategies (Table \ref{tb:rvins}).  HARPS-North, installed on the
Telescopio Nazionale Galileo (TNG) on La Palma Island in the Canary
Islands \citep{Cosentino12}, is based on the design of the original
HARPS instrument but will use a laser comb to achieve a precision of
$\sim$10 cm s$^{-1}$ \citep{Li12}.  The Echelle SPectrograph for Rocky
Exoplanet and Stable Spectroscopic Observations (ESPRESSO) on the ESO
VLT is expected to reach an RV precision of at least $10$~cm~s$^{-1}$,
with a goal of a few cm~s$^{-1}$ \citep{Pepe10}.  The design
specifications of CODEX, planned for the European Extremely Large
Telescope, call for $<$2 cm s$^{-1}$ RV precision \citep{Pasquini10}.
If there is an equivalent suppression in the effect of stellar
``jitter'', such instruments should be able to find STEPs orbiting
close ($P < 10$~d) to nearby bright stars
(Fig. \ref{fig:detectability}). However, estmating the yield of a
survey is difficult because of the lack of data on stellar noise at
such precision, and thus any empirical means to construct a suitable
target catalog of Doppler ``quiet'' stars.  However, should an
instrument such as CODEX achieve $\sim2$ cm~s$^{-1}$ precision it
could, in principle, detect Mars-size planets with $P < 100$~d
(Fig. \ref{fig:detectability}).
    
\begin{table}[h]
  \centering
  \begin{minipage}{130mm}
    \caption{Current and Future Spectrographs for Radial Velocity Measurements.}
    \label{tb:rvins}
    \begin{tabular}{lrllrl}
      \hline\noalign{\smallskip}
      Instrument & Precision & First Light & $\lambda$ & Resolution & Reference\\
      & (cm~s$^{-1}$) & & ($\mu$m) & & \\
      \hline
      Keck-HiRES & 100 & 1996 & 0.3--1.0 & 85,000 & \citet{Butler96} \\
      UCLES & 300 & 1998 & 0.48--0.86 & 45,000 & \citet{Butler01}\\
      HDS & 300 & 2000 & 0.35--0.65 & 150,000 & \citet{Kambe08}\\
      HARPS & 50 & 2003 & 0.38--0.69 & 115,000 & \citet{Rupprecht04}\\
      SOPHIE & 200 & 2006 & 0.38--0.69 & 75,000 & \citet{Bouchy13}\\
      CHRONIS & 70\footnote{Expected performance} & 2011 & 0.45--0.89 & 120,000 & \citet{Schwab12}\\
      HARPS-N & 50$^a$ & 2012 & 0.38--0.69 & 115,000 & \citet{Cosentino12} \\ 
      APF-Levy & $\sim$200$^a$ & 2013\footnote{In commissioning phase} & 0.3--0.65 & 73,000 & \citet{Radovan10} \\
      APF-HWS & $\sim$100$^a$ & 2014 & 0.38--0.69 & 100,000 & A. Howard, priv. comm. \\ 
      ESPRESSO & $\sim$10$^a$ & 2016 & 0.35--0.72 & 150,000 & \citet{Pepe10} \\ 
      CODEX & $\sim$2$^a$ & 2025 & 0.37--0.72 & 150,000 & \citet{Pasquini10} \\ 
    \end{tabular} 
  \end{minipage}
\end{table}
    
\section{Other Detection Techniques and Sub-Earth Objects}
\label{sec:other}
\subsection{Pulsar planets}

Since the discovery of the three planets of pulsar PSR1257+12,
including one of lunar mass, searches of several dozen other
millisecond pulars have revealed no other systems of similar ilk
\citep{Wolszczan2012}.  Pulsar PSR1719-1328 has a single substellar
($\sim$2--3 $M_J$) companion on a 2.2~hr orbit \citep{Bailes2011}, but
this may be the degenerate helium or carbon/oxygen remnant of a former
``donor star" \citep{vanHaaften2012}.  Given the exquisite timing
stability of millisecond pulsars, the lack of additional discoveries
cannot be an artifact of sensitivity.  \citet{Miller2001} propose that
the scarcity of planets around millisecond pulsars can be explained in
terms of the ``recycling" hypothesis where accretion of matter from a
donor star spins up the pulsar and makes it emit extremely stable,
detectable radio signals.  Such accretion produces an X-ray luminosity
sufficient to vaporize any planets, and \citet{Miller2001} argue that
PSR1257+12 must be a rare example of a high primordial spin.  Even
more problematic is developing a plausible mechanism for the formation
of the planets, either via survival of the supernova explosion that
created the neutron star, by accretion from the disk resulting from
the disruption of a companion or supernova fallback, or by capture
from a main-sequence star \citep{Sigurdsson1993}.  We refer the reader
to the review by \citet{Phillips1994} and references therein, as well
as the revisit by \citet{Hansen2009} to this issue.  Although
additional pulsar planets may be uncovered in the future, their rarity
means that they will not significantly contribute to the catalog of
known sub-Earths.
    
\subsection{Microlensing}
    
A small planet can also be detected by ``microlensing", i.e. as its
host star passes very close to the line of sight between an observer
and a more distant star \citep{Mao1991}.  The effect of the planet is
to break the radial symmetry of the gravitational lens and produce a
distinctive, hours-long deviation from the symmetric days-long
amplification in the light curve of the background star.  Microlensing
events are rare and this technique requires simultaneous monitoring of
millions of distant stars, e.g. in the Galactic Bulge.
    
The method is most sensitive to planets with semi-major axes of a few
AU, i.e. the angular Einstein radius projected to the typical distance
of a lens.  Sensitivity to small planets is ultimately limited by the
angular size of the background star compared to the Einstein radius of
the planet.  In principle, monitoring of giant stars in the Bulge
permits the detection of planets as low as 1~\mearth{}
\citep{Bennett1996}, but because of limited observing cadence and
sensitivity, the smallest planet detected to date has a mass of a few
Earths \citep{Kubas2012}. Routine detection of Earth-size planets will
require a second generation of ground-based microlensing surveys
\citep{Wright2012}.
    
For dwarf stars at the distance of the Bulge, the theoretical
detection limit is a few lunar masses.  However, these stars will be
very faint ($V > 20$), and the projected surface density of such stars
towards the Bulge is several per square arc-second.  Therefore, a
dedicated space telescope that can achieve diffraction-limited, high
photometric precision observations is required
\citep{Bennett2002,Bennett2008}. Such aspirations may eventually be
realized in the form of the \emph{Euclid} mission \citep{Penny12}, the
Wide Field InfraRed Space Telescope (WFIRST) \citep{Barry2011}
mission, or the NEW-WFIRST mission \citep{Dressler2012}, which should
be capable of detecting planets as small as Mars ($\sim$0.1~\mearth{}).
In particular, two 2.4 m telescopes, built by the US National
Reconnaissance Office and transfered to NASA offer diffraction limited
imaging of 0.16 arc-seconds at $\lambda = 1.6$~$\mu$m and might triple
the yield of Mars-size planets compared to the design reference
mission of WFIRST \citep{Dressler2012}.
    
\subsection{Exomoons \label{subsec:exomoons}}
    
Satellites of exoplanets have yet to be discovered but the Copernican
idea predicts that the Solar System is not unique in this way.  Given
that Ganymede, the largest satellite in our Solar System, has only
2.5\% of Earth's mass, it seems likely that satellites in other Solar
Systems will also be sub-Earth objects.  Satellites can accrete from
the circumplanetary disks of giant planets, which accrete inflowing
gas and solid material from the circumstellar disk.  Dynamical
simulations suggest that if a satellite grows large enough it will
migrate inward and be accreted by its host planet
\citep[e.g.,][]{Canup06,Sasaki10}.  \citet{Canup06} found that this
process restricts the ratio of cumulative satellite mass to planet
mass to $\lesssim 10^{-4}$ e.g. $\sim 0.1$~\mearth{} for a
Jupiter-mass planet.  This limit depends weakly on model parameters
including gas surface density, abundance of solids, inflow timescale
and migration efficiency.  If the \citet{Canup06} model is
representative of typical circumplanetary disk accretion, we should
not expect satellites formed in-situ around Jupiter-mass planets to
greatly exceed the mass of Mars. On the other hand, Ogihara et
al. (2012) added an inner circumplanetary disk cavity to the model of
\citet{Canup06}, and found that, in some cases, inward migration was
inhibited.  An inner cavity can be caused by magnetic coupling of the
planet to the disk, but it is unclear how frequently this occurs.
    
Not all exomoons would necessarily form in a circumplanetary
disk. The irregular orbits of many Solar System satellites indicate
that they were acquired by their giant planet hosts through one of
several proposed capture mechanisms.  Dynamical simulations by
\citet{Cuk04} suggest that if a planetesimal encountering a
circumplanetary disk experiences sufficient gas drag, capture could
occur.  Alternatively, binary planetesimals could become separated by
the tidal pull of a giant planet resulting in the capture of a single
planetesimal. In fact, the inclined retrograde orbit of Neptune's moon
Triton might be the result of the latter mechanism \citep{Agnor06}.
Even if capture events in exoplanet systems are rare, the rate of
survival of captured moons might be high. \citet{Porter11} considered
the post-capture stability of satellites around non-migrating giant
planets in stellar habitable zones and found that about 20--50\% of
Mars- to Earth-size satellites enter stable orbits. It is conceivable
that terrestrial-size planets could also be captured by an inward
migrating gas giant. However, simulations are needed to estimate the
likelihood of such events as well as the likelihood of long-term
orbital stability.  As a gas giant moves closer to its host star, its
Hill radius shrinks and stability is limited to tighter orbits.

Close encounters between giant planets might commonly eject their
satellites: \citet{Gong13} performed dynamical simulations consisting
of three satellite-hosting giant planets, with varying planet mass,
planet eccentricity, satellite mass, and satellite semi-major axis.
In $\sim$75\% of their simulations, planet-planet scattering resulted
in the ejection of all satellites.  If planet-planet scattering is
common among giant planets they are unlikely to retain their
primordial satellites.  The prospect for satellites around hot Jupiters
would be especially grim if the orbits of these planets is a result of
strong dynamical interactions.
    
Moon-size or larger satellites could of course form in the manner of
the Moon's proposed origin --- from the collision of two
(proto)planets.  \citet{Elser11} predict that satellites with $>$0.5
lunar masses form around approximately 1 in 12 rocky planets. The most
massive satellite produced in their simulations has three lunar
masses.
        
In principle, \emph{Kepler} can discover massive exomoons in the same
manner it finds sub-Earths.  An exomoon can transit the host star,
adding an additional transit signal differing in phase, amplitude, and
duration from that of the planet.  Satellites on wide orbits could induce
detectable reflex motions in the planet \citep{Kipping09}, while
satellites on close-in orbits are more likely to transit the planet as
the planet transits the star. The latter is called a ``mutual event"
during which the transit signal temporarily \emph{decreases}
\citep{Ragozzine10, Pal12}.  Satellites with large orbital
inclinations relative to their host planet's orbital plane could
transit the star even if the planet does not.  

Satellites can also be detected by TTVs as well as the transit
duration variations (TDVs) that they induce on their planetary hosts.
The TTV signal is shifted in phase by $\pi/2$ relative to the TDV
signal, allowing it to be distinguished from the effect of another
planet \citep{Kipping09}.  Moreover, TTV amplitudes scale differently
with satellite mass and orbital distance than TDV amplitudes, and
these complementary measures thus provide unique solutions for these
two parameters.  Mass-dependent TTV and TDV detections can be combined
with a (radius-dependent) transit signal to calculate the satellite's
bulk density and infer its composition.  \citet{Kipping09} found that,
using TTV and TDV measurements alone, \emph{Kepler} may be sensitive
to exomoons $\ge$ 0.2~\mearth{} or about 0.67~\rearth{} for an
Earth-like composition.  Hence, \emph{Kepler} should be able to detect
both the transit (\S \ref{subsec:transit}) and the TTV/TDV signals
from such a massive exomoon --- if they exist.
    
The Hunt for Exomoons with Kepler (HEK) has identified the most likely
hosts of detectable exomoons based on \emph{Kepler} light curves
\citep{Kipping13}.  The light curves of selected stars are compared
with planet-only or planet-satellite models in a Bayesian analysis. So
far, the HEK team has shown that the 7 most likely candidates are
unlikely to have moons comparable in mass to their 2--4~\rearth{}
planet hosts \citep{Kipping13}.
    
\section{Formation of Subterrestrial Planets}
\label{sec:formation}
    
According to canonical theory, rocky planets accrete from a disk of
gas and dust that surrounds a protostar.  In the first $\sim10^{4}$
years, micron-size dust grains coagulate and settle to the disk
midplane.  However, growth of particles to sizes larger than a few
millimeters cannot be observed and is poorly understood.  Laboratory
experiments indicate that collisions of millimeter-size grains rarely
lead to sticking and growth under the assumed dynamical conditions
within disks \citep{Blum00, Zsom10, Weidling12}.  Moreover, the motion
of larger (meter-size) bodies decouples from the gas and they
experience a headwind and orbital decay into the star
\citep{Weidenschilling77b}.  In theory, meter-size bodies should be
lost to the central star in $\sim$100 years \citep{Youdin12}.
Possible mechanisms by which nature overcomes these growth barriers
have been proposed \citep[See][for reviews]{Chiang10, Morbidelli12,
  Youdin12, Haghighipour13}.  Regardless, bodies large enough
(km-size) to be impervious to this effect must form by some mechanism.
Once mutual gravitation begins to dominate, larger objects experience
runaway growth over $10^{5}-10^{6}$ years \citep{Morbidelli12}.
Growth of the largest bodies slows down as they accumulate most of the
remaining material within a ``feeding zone", becoming lunar-to
martian-size protoplanets \citep{Chambers06, Kokubo98, Kokubo00}.
Once the mass in protoplanets exceeds that of planetesimals, their
orbits begin to cross \citep{Kenyon06}.  Chaotic scattering and
collisions ensue for ~$\sim$100 Myr until a few relatively isolated
planets remain.

STEPS might emerge from the final stages of planet formation via
the same mechanisms proposed for their Earth- and super-Earth-size 
brethren: (i) as the in-situ products of constructive and destructive 
collisions and scattering of smaller planetary embryos 
\citep{Morbidelli12, Kennedy08}; (ii) from embryos that have migrated
inwards as a result of gravitational torques exerted by the
protoplanetary disk \citep[``Type-I migration",][]{Ida10,Terquem07,O'Brien06}, (iii) by gravitational scattering
\citep{Ida10, Kennedy08, Raymond08}, or (iv) shepherding by 
inward-migrating giant planets \citep{Zhou05,Raymond06,Fogg07,Mandell07} or
super-Earths \citep{Kennedy08}.  A fifth mechanism - evaporation of
larger bodies \citep{Valencia10}, is discussed in \S
\ref{sec:properties}.  The rate of Type-I migration of an object is
proportional to its mass \citep{Ward97} whereas smaller planets are
more susceptible to gravitational scattering and shepherding.  For a
more detailed review of rocky planet formation see
\citet{Morbidelli12}.
    
Numerical N-body simulations are a popular tool for investigating the
late stages of rocky planet formation \citep{Morbidelli12}. Such
simulations consistently produce planets that have masses between
those of Mars and Venus over a wide
range of orbits (e.g. \citet{Walsh11, Raymond09, Montgomery09,
  Kokubo06}). These outcomes are plausible examples of other planetary
systems produced by the stochastic nature of the planet formation
process, suggesting that most STEPs are unlikely to be of similar mass
or occupy similar orbits as Mercury or Mars.
    
\begin{figure}[h]
  \centering
  \includegraphics[scale=0.32]{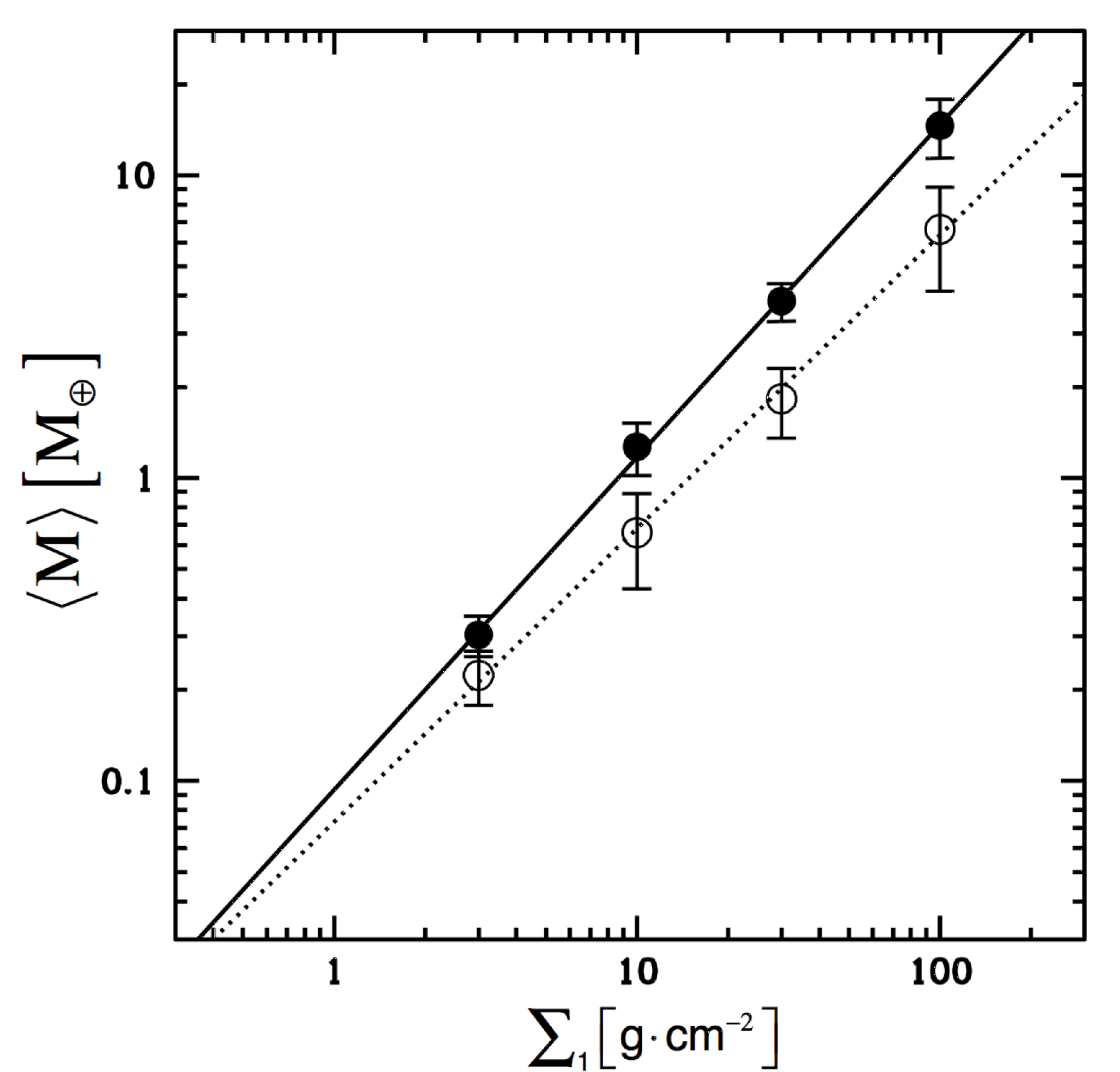}
  \caption{Mean masses $\left<M\right>$ of the largest (filled
    circles) and second largest (open circles) planets formed in
    simulations of rocky planet formation as a function of disk
    surface density \citep{Kokubo06}.  The disk surface density
    depends on radius as a power-law with index -3/2, and is
    $\Sigma_1$ at 1~AU.  Best-fit power law relations between
    $\Sigma_1$ and $\left<M\right>$ have indices of 1.1 and 0.97
    (solid and dashed lines, respectively).}
  \label{fig:M_vs_Rho}
\end{figure}

The initial mass surface density of the disk is an important parameter
of dynamical models and it may determine planet size. One common
choice of initial condition is the Minimum Mass Solar Nebula
\citep[MMSN;][]{Weidenschilling77,Hayashi81}. However, the MMSN is not
necessarily representative of all disks.  For example,
\citet{Chiang12} derive a ``Minimum-Mass Extrasolar Nebula" from the
population of \emph{Kepler}-detected super-Earths that is $\sim$5
times denser than the MMSN.  Figure \ref{fig:M_vs_Rho} shows the
average mass of the largest and second-largest planets that form between 0.5 and 1.5 AU in
simulations by \citet{Kokubo06}.  The authors varied the mass surface density
of the disk at 1~AU ($\Sigma_{1}$) while maintaining a power-law radial surface density profile
with index -3/2.  The MMSN corresponds to
$\Sigma_{1}=7$~g~cm$^{-2}$. The average masses follow
power laws with disk surface density having indices close to unity. If
low-mass stars have disks of lower surface density, then the results
of \citet{Kokubo06} predict that small planets are most common around
M-dwarfs, at least at 0.5--1.5 AU. However, the evidence for a
relation between disk surface density and stellar mass is tentative
\citep{Williams11,Andrews13}.
    
The limited supply of disk material and hence the surface density
likely governs the in-situ formation of close-in and especially
detectable planets. \citet{Raymond07} show that a MMSN disk rarely
forms planets more massive than Mars within $\simeq$0.1 AU of late
M-dwarfs. However, \citet{Montgomery09} find that a disk three times
denser than the MMSN instead produces 3-5 planets with an average mass
of 0.7--0.8~\mearth{}. These results suggest that the formation of STEPs
depends on surface density. However, \emph{Kepler} observations show
that the occurrence of Earth- to Neptune-size planets does not depend
on stellar metallicity, a potential proxy for disk surface density
\citep{Buchhave12,Mann13}.
    
STEPs may also migrate to close-in orbits after forming at larger
orbital distances. \citet{Kennedy08} conclude that the minimum mass of
a planet able to migrate to a short-period orbit is proportional to
the distance at which ice condenses in the disk (the ``snow
line"). This distance scales with stellar luminosity and hence stellar
mass. The same study shows that planets also tend to scatter to
smaller orbital distance around low-mass stars, again suggesting that
low-mass stars might commonly host detectable STEPS.  Because (Type I)
orbital migration scales with planet mass and mutual gravitational
interactions will preferentially scatter less massive planets, we
expect scattering to be more efficient than migration in dispersing
the orbits of STEPS.
    
\section{Properties of Subterrestrial Planets}
\label{sec:properties}
    
Although our Solar System lacks objects with radii between that of
Mars and Venus, the diversity among smaller bodies (e.g., Ganymede,
Titan, Mercury, and Mars) suggests that STEPs might have diverse
characteristics depending on composition, distance from the star, and
contingencies such as giant impacts.  While there are many properties
of such bodies that are of interest, we focus on two that may be
ascertainable in the near future: mean density, and the presence or
absence of an atmosphere.
    
The mean density $\bar{\rho}$ of a planet can be estimated if both
radius and mass are measured (by the transit and Doppler methods,
respectively).  These can be compared with theoretical mass-radius
relations to infer composition
\citep[e.g.][]{Valencia07,Seager07,Grasset09,Rogers10}.  Although such
interior models have not been applied to STEPs per se, there have been
detailed comparisons of similar models with Mercury \citep{Hauck13}
and Mars \citep{Sohl05}.  Comparison with STEPS will be limited
by Doppler precision (discussed in \S \ref{sec:rv}), sensitivity to
planet radius and hence stellar radius \citep[$\ge$3\%][]{Torres10},
and degeneracies between composition and mean density
\citep{Rogers10}.  Because of these limitations, observations will
likely be able to discriminate only between the most extreme
compositions, i.e. very volatile-rich, rocky, or metal-rich planets.
    
Mercury, as the Solar System's smallest and innermost planet, is
arguably the most appropriate analog to those sub-Earths that can be
detected by the transit (\S \ref{sec:kepler}) and Doppler (\S
\ref{sec:rv}) methods.  Mercury's salient bulk properties are a
comparatively large iron core comprising most of its volume and mass
\citep{Hauck13}, and a lack of volatiles or substantial atmosphere.
Whether its oversized core reflects a non-chondritic composition for
the primordial disk close to the Sun \citep{Lewis72,Weidenschilling78,
  Ebel12}, or is a consequence of removal of most of the silicate
mantle by a giant impact \citep{Benz88,Benz07,Gladman09} is debated.
Both of these mechanisms are more effective on orbits closer than
Mercury ($P = 88$~d), where temperatures in a planet-forming disk and
the kinetic energy of impacts, which scales with orbital velocity, are
higher.  This implies that STEPs close to their host stars may have
comparatively large cores as well.
    
Like Mercury, STEPs on close-in orbits are likely to lack any
substantial atmosphere because of their weak gravity and heating of
their upper atmospheres by stellar X-ray, extreme ultraviolet (EUV)
and Lyman-$\alpha$ radiation.  Rapid atmospheric escape is predicted
for their more massive super-Earth counterparts \citep{Tian09,
  Pierrehumbert11}.  In the limit that the thermal speeds of atoms
become comparable to escape speeds, hydrodynamic escape ensues and
mass loss rate is limited only by the rate at which energy is absorbed
by the atmosphere:
\begin{equation}
  \label{eqn.escape}
  \dot{M} = \frac{3\epsilon F}{4G\bar{\rho}},
\end{equation}
where $F$ is the incident flux absorbed by the atmosphere, $G$ is the
gravitational constant, and $\epsilon$ is an efficiency factor that
accounts for the inflation of the atmosphere and radiative,
conductive, and evaporative cooling.  In this regime, absorption of
energy $E$ per unit area results in a loss of atmosphere (in units of
pressure) of $E/(4\pi R_p)$.  Given realistic models of the evolution
of the X-ray and UV output of dwarf stars
\citep{Ribas05,SanzForcada11}, a Mercury-size planet on a 10-day orbit
around its host star is expected to lose thousands of bars of
atmosphere over billions of years.
    
Hydrodynamic escape will occur from the top of the atmosphere (the
exobase) only if it is hot enough and the Jeans parameter $\lambda =
GM_p\mu/(R_ek_BT)$, the ratio of the gravitational potential to the
thermal energy, is $<$2.8 \citep{Volkov10}, where $\mu$ is the atomic
mass, $R_e$ is the distance of the exobase from the planet's center,
and $k_B$ is the Boltzmann constant.  As the thermal energy approaches
the escape energy, the atmosphere inflates, $R_e > R_p$, and the escape
energy decreases.  Approximating this inflation as $R_e \approx R_p +
h$, where $h = k_BTR_p^2/(GM_p\mu)$ is the atmospheric scale height,
the required exobase temperature is $\sim$0.29$GM_p\mu/(R_pk_B)$.
Under all plausible conditions, the light elements H and He will
hydrodynamically escape from close-in planets, carrying some heavier
elements with them.
    
\begin{figure}[h]
  \centering
  \includegraphics[scale=0.45]{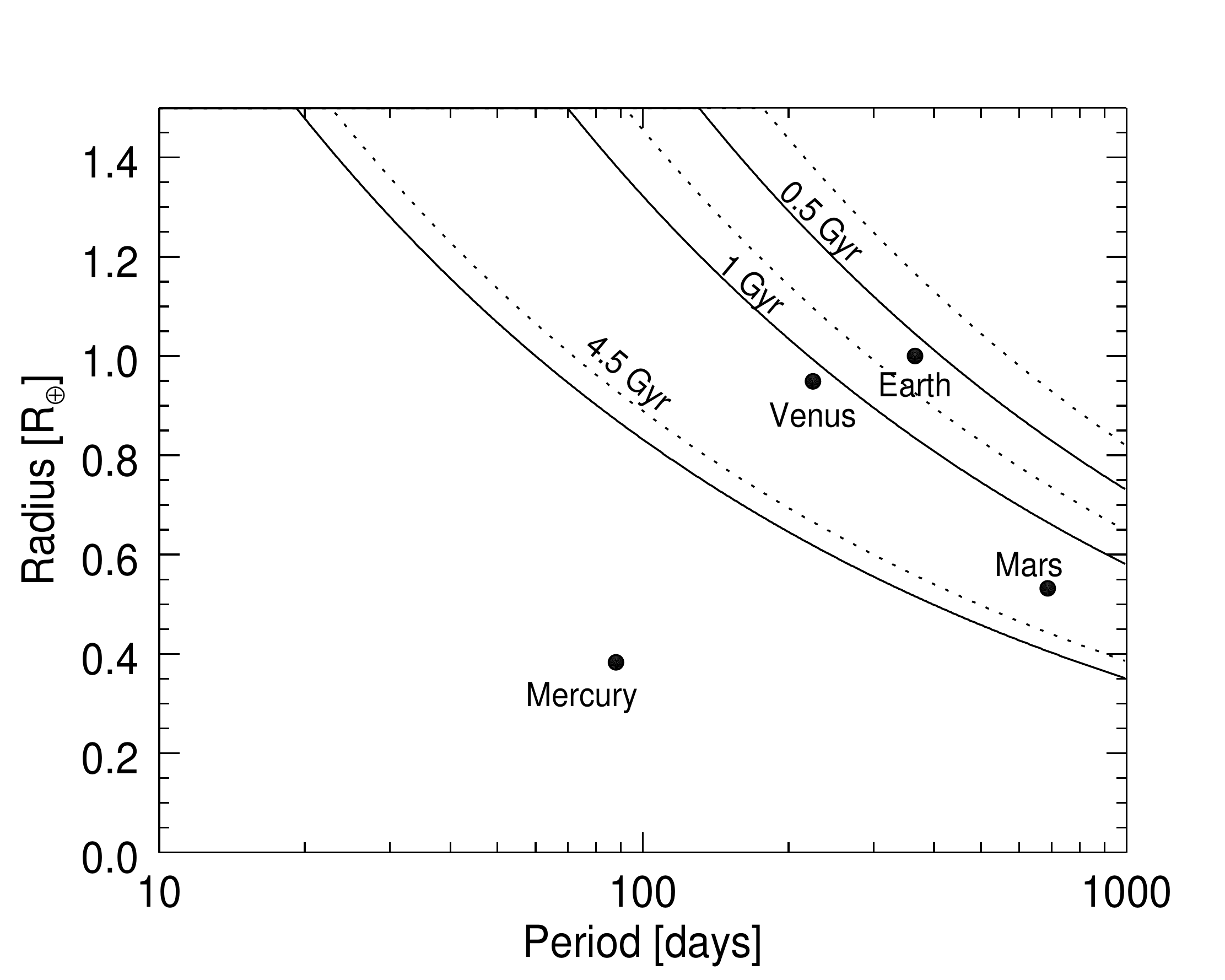}
  \caption{Combinations of rocky planet radius and orbital period for
    which the Jeans parameter $\lambda$ of a CO$_2$ atmosphere is 2.8,
    the condition for hydrodynamic escape.  The combined fluxes of
    stellar X-ray, EUV, and Lyman-$\alpha$ radiation at ages of 0.5,
    1, and 4.5~Gyr are used \citep{Ribas05,SanzForcada11}.  Solid
    lines are for a Sun-like star and dashed lines are for an M0 dwarf
    star host.  Atmospheres below and to the left of these lines will
    have $\lambda < 2.8$ and be hydrodynamically escaping.  The inner
    planets of the Solar System are plotted.}
  \label{fig:escape}
\end{figure}

A more germane question is whether hydrodynamic escape continues when
H and He are exhausted, e.g., by the atomic oxygen produced by
dissociation in a CO$_2$-dominated Venus-like atmosphere.  For atomic
oxygen on Venus, the required temperature is $3 \times 10^4$~K, but
only $\sim$5000~K on a planet the size of Mercury.  Figure
\ref{fig:escape} plots the $\lambda = 2.8$ condition for planets with
CO$_2$ atmospheres and the expected combination of X-ray, EUV and
Lyman-$\alpha$ irradiation \citep{Ribas05,SanzForcada11} for Sun-like
(solid lines) and M0 dwarf (dashed lines) stars at three stellar ages.
Hydrodynamic escape ($\lambda < 2.8$) occurs to the lower left of each
boundary.  These calculations assume that all incident X-ray plus UV
(XUV) energy is absorbed at the top of the atmosphere and conducted
downward, principally by atomic oxygen, to the homopause, then
radiated away in the infrared by CO$_2$.  We follow the procedure in
\citet{Pierrehumbert11}, except that by assuming a constant thermal
conductivity $k$, an analytical solution is available for the required
irradiation $q$ as a function of $\lambda$:
\begin{equation}
  \label{eqn.irradiation}
  q = \frac{G M_p \mu k}{R_p^2k_b}\ln \frac{\sigma R_p p_0}{\lambda k_B T_0} \left[\ln \frac{GM_p\mu}{\lambda R_p k_B T_0}\right]^{-1},
\end{equation}
where $p_0$ and $T_0$ are the pressure and temperature at the
homopause, and $\sigma$ is the collision cross-section.  We calculate
the $k$ of atomic O using \cite{Dalgarno62}, assume a homopause
pressure like that of Venus ($10^{-3}$ Pa) and homopause temperature
equal to the planet's equilibrium temperature, and adopt $\sigma = 2
\times 10^{-19}$~m$^{2}$ \citep{Tully01}.  Equation
\ref{eqn.irradiation} approximately reproduces the Jeans parameter for
O on current Venus ($\lambda = 260$) and predicts that it lost
atmosphere by hydrodynamic escape prior to 3.5~Ga (if indeed it had a
CO$_2$ atmosphere).  It also predicts the hydrodynamic escape of
CO$_2$ from Mars in the past.  Mercury would have suffered
hydrodynamic escape of any CO$_2$ atmosphere throughout its history.
Any sub-Earth on a closer orbit would have experienced yet greater
loss.  This is in addition to any removal by the stellar wind
\citep{Zendejas10} or impacts \citep{Ahrens93}.
    
The presence or absence of a substantial atmosphere might be
discernable by follow-up observations, at least if the planet is very
close to its star, very hot, and tidally locked.  By detecting the
infrared emission from the planet and measuring its variation with
phase, the redistribution of heat around a synchronously rotating
planet can be estimated
\citep[e.g.,][]{Gaidos04,Lewis10,Cowan11,Demory12}.  Planets lacking
an atmosphere will have no redistribution, their substellar
hemispheres will be hotter, and their phase curves more pronounced.
Planets with a thick, circulating atmosphere will have cooler
illuminated hemispheres because some of the heat is transferred to the
night side, and their emission will exhibit little or no variation
with phase.  The boundary between these two regimes has not been
theoretically established for planets on close-in orbits but is
probably equivalent to a surface pressure of a fraction of a bar.  The
emission from an unresolved transiting planet can be detected by
differencing the signal in and out of secondary eclipse.  In
exceptional cases, it might be possible to determine the planet's
phase curve by measuring the small variation in total flux over a
complete orbit.
    
\begin{figure}[h]
  \centering
  \includegraphics[scale=0.3]{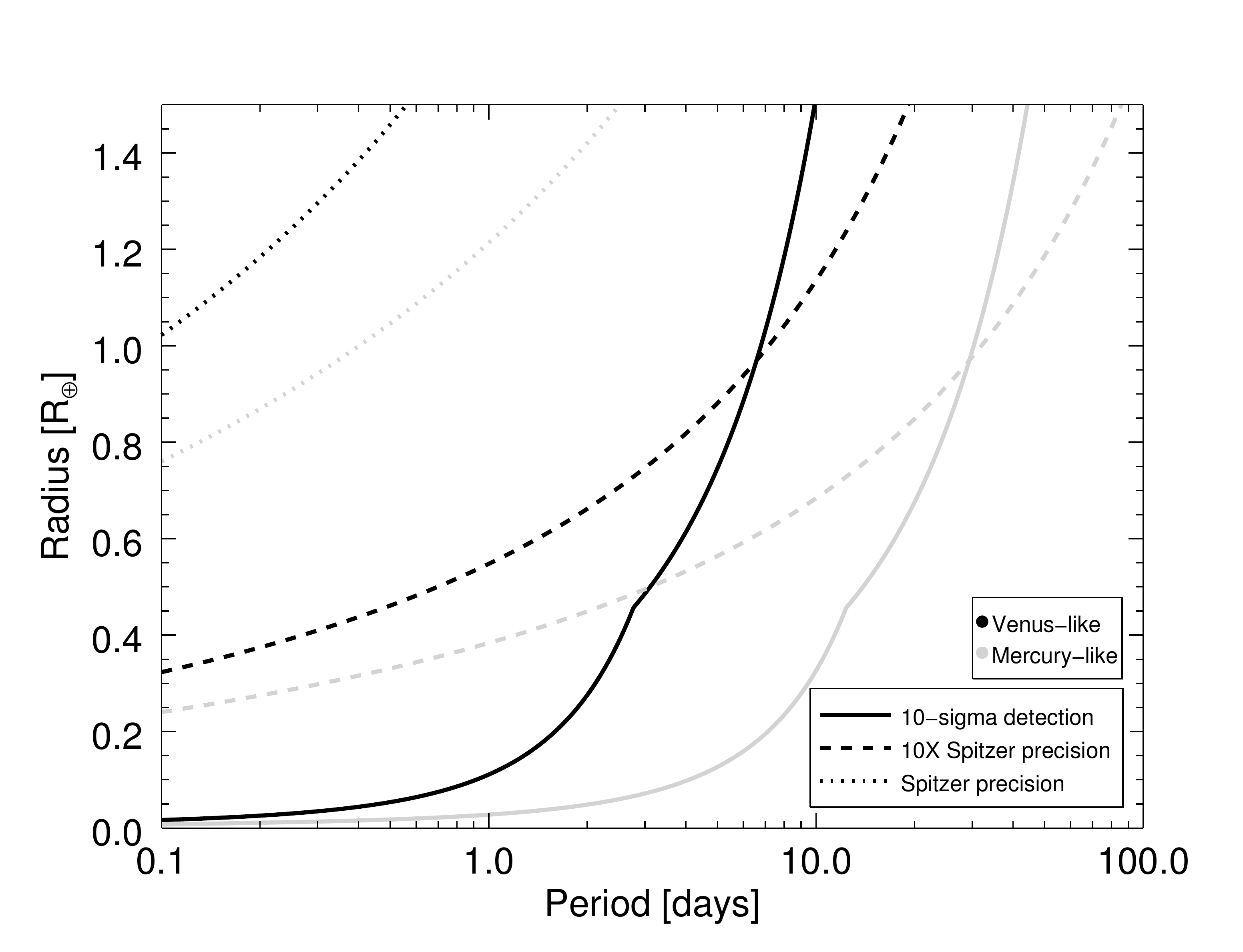}
  \caption{Detection by differential photometry of an eclipsing STEP
    around a M0 dwarf star at 10~pc with the \emph{James Webb} Space
    Telescope and either MIRI or NIRCam.  Two cases are considered: a
    Venus-like albedo and efficient re-distribution of heat around the
    planet (black lines), and a Mercury-like albedo, no redistribution
    of heat, and isotropic emission (grey lines).  The solid lines are
    the $10\sigma$ detection of an isolated source, while the dotted
    and dashed lines are the detection limits if the stellar signal is
    removed with a photometric accuracy of $10^{-4}$ (typical of
    \emph{Spitzer} observations) or $10^{-5}$.  The actual stability
    of these instruments will not be known until JWST is launched.}
  \label{fig:jwst}
\end{figure}

The most promising (and perhaps only) tool with which to carry out
such observations will be JWST using either the Near Infrared Camera
(NIRCam) or the Mid Infrared Instrument (MIRI) \citep{Clampin12}.
Figure \ref{fig:jwst} shows estimated detection thresholds vs. orbital
period for sub-Earths orbiting an M0 dwarf star at 10~pc.  Breaks in
the curves mark transitions between the regimes where one instrument
is favored over the other.  Two cases are considered: a Venus-like
planet with an albedo of 0.9 and efficient heat redistribution (black
curves), and a Mercury-like planet with an albedo of 0.068 and no heat
redistribution (grey curves).  For each case, we calculate a
$10\sigma$ detection threshold in terms of the minimum angular radius
of the planet, i.e. its physical radius at a distance of 10~pc (solid
curves).  We assume blackbody emission, a $10^4$~s integration and the
sensitivities from the JWST
website\footnote{http://www.stsci.edu/jwst/science/sensitivity/jwst-phot}.
In principle, JWST can detect the thermal emission from very small hot
planets.  However a more relevant measure of sensitivity is the
accuracy with which the stellar signal can be subtracted, i.e. the
photometric stability.  The dotted curves in Fig. \ref{fig:jwst}
correspond to a fractional detection threshold of $10^{-4}$ relative
to the host star.  This level of stability has been achieved with the
\emph{Spitzer} infrared space telescope \citep{Demory12}.  The two
dashed curves are for a hypothetical stability of $10^{-5}$ (the
actual stability will not be known until JWST is in space).  It
appears that sub-Earths can be detected by JWST only if its stability
significantly exceeds that of \emph{Spitzer} and only if the planets
lack substantial atmospheres.
    
\clearpage
    
Removal of mass from sub-Earths may not stop with a CO$_2$ atmosphere.
On close-in, tidally-locked planets, substellar surface temperatures
are $T \approx 2800(L_*/L_{\odot})^{1/4}(P/{\rm 1d})^{-1/3}$ where
$L_*$ is the stellar luminosity and a weak dependence on stellar mass
is ignored, and a Mercurian albedo (0.068) is assumed.  If
temperatures exceed the melting point of silicates, a magma ``sea"
would be present \citep{Leger11}, and a tenous silicate vapor (SiO, O,
and Si) atmosphere would form \citep{Miguel11}.  Continuous escape of
this atmosphere would cause gradual mass
loss\footnote{\citet{Cameron85} and \citet{Fegley87} proposed that the
  mantle of Mercury was evaporated, but this is inconsistent with the
  MESSENGER estimate of volatile radioactive potassium in the
  Mercurian crust \citep{Peplowski11}.}.  \citet{Valencia10} assumed
$\epsilon \approx 0.4$ in Eqn. \ref{eqn.escape} and estimated that the
hot ``super-Earth'' CoRoT-7b ($P = 0.85$~d) may have lost about half
of its mass.  This phenomenon would occur even more readily on smaller
planets.  Extreme rates of evaporation and a coma of silicate
condensates have been proposed to explain the variable transit depth
of whatever object orbits the {\it Kepler} star 12557548 with a period
of 0.65~d \citep{Rappaport12}.
    
In this scenario, the silicate mantles of sub-Earths close to
solar-type stars may largely evaporate.  As evaporation proceeds, the
residual mantle, mixed by melting, would become steadily enriched in
more refractory, heavier elements.  It could eventually founder and/or
dissolve into the core, whereupon the object would become an ``iron
planet'', a naked core with a relatively high mean density.  Without a
comparatively light element such as O, hydrodynamic escape would halt,
although the stellar wind might continue to erode an iron vapor
atmosphere.  This scenario would not unfold around M dwarfs with $L_*$
as low as $10^{-4}L_{\odot}$; the equilibrium temperatures of
sub-Earths around such stars would be up to $10\times$ cooler and they
would retain their silicate mantles.
    
\section{Discussion}
\label{sec:discussion}

Over the past two decades, successive discoveries enabled by
improvements in ground-based instruments and space missions such as
\emph{Hubble}, \emph{Spitzer}, CoRoT, and \emph{Kepler} have uncovered
brown dwarfs, Jupiter-like gas giants, volatile-rich Neptunes,
``super-Earths'', and now Earth-size and presumably rocky planets.
The discovery and characterization of sub-Earths, planets with masses
significantly less than that of Earth, is the next and perhaps
ultimate leg of the scientific journey to enumerate the worlds on
close-in orbits around main-sequence stars.  This step has just begun,
but we can already draw the following conclusions:
    
\begin{itemize}
    
\item Several dozen confirmed or candidate sub-Earths have already
  been discovered by \emph{Kepler} (plus 3 others by the Arecibo
  telescope and \emph{Spitzer}).  We expect that number to grow as the
  remaining \emph{Kepler} data is analyzed.  Studies of \emph{Kepler}
  transit light curves may also reveal exomoons, if sufficiently
  massive ones exist.\\
    
\item Enumeration of \emph{Kepler} sub-Earths will determine whether
  the planet distribution with radius remains flat \citep{Fressin13},
  rises, or falls below $\sim$1~\rearth{} \citep{Petigura13}.  Based on
  the number of \emph{Kepler} discoveries to date ($\sim$40), and a
  detection efficiency of 5\% among the $\sim$33,000 most suitable
  stars (Fig. \ref{fig:detectprob}), we estimate that at least 2--3\%
  of stars have planets with 0.5~\rearth{} $< R_p < 1$~\rearth{} and
  $P < 10$~d.\\
    
\item Sub-Earths are at the limit of \emph{Kepler}'s detection
  threshold and any estimate of their occurrence is sensitive to
  completeness for very small signals, which is still being determined
  \citep{Petigura13}.  Moreover, the estimated radii of transiting
  planets depend on the radii of the host stars; those of
  \emph{Kepler} targets are being refined as stellar parameters are
  measured and improved models are applied \citep{Muirhead12b,Mann12}.
  As a result, our estimate of 2--3\% should be considered very
  tentative and probably a lower limit.\\
    
\item With forseeable instruments, only STEPs with orbital periods of
  a few days will be detectable by Doppler
  (Fig. \ref{fig:detectability}).  \emph{Kepler} stars are too faint
  for observations of such precision, but it is conceivable that a
  sample of nearby, much brighter stars could be interrogated by the
  Doppler method and the two populations compared by statistical
  techniques \citep[e.g.,][]{Gaidos2012}.  The occurrence of close-in
  STEPs could be compared with the number of such objects detected on
  wider orbits by a microlensing mission such as WFIRST or NEW-FIRST.
  Such a study would investigate whether sub-Earths preferentially
  form on close-orbits, or are dynamically scattered onto distant
  orbits by their more massive counterparts.\\
    
\item Planet formation theory predicts that the surface density of
  disks influences the size of planets that form in-situ.  This
  linkage may be difficult to reconcile with the apparent independence
  of small (Earth- to Neptune-size) planet occurrence and host star metallicity,
  unless migration is common.  Determining whether or not these
  parameters remain independent in the sub-Earth regime, where orbital
  migration is expected to be less efficient, will help resolve this
  apparent discrepancy.\\
    
\item If the mass or surface density of planet-forming disks scales
  with that of the star, then M dwarfs may preferentially host
  sub-Earths, boding well for their detection
  (Fig. \ref{fig:detectability}).  However, this premise is only
  weakly supported by the available data and the occurrence of small
  planets may not depend on stellar mass \citep{Fressin13}, as was
  originally thought \citep{Howard12}. The Atacama Large Millimeter
  Array (ALMA) will help to clarify any relation between surface
  density and stellar mass in the context of planet formation.  Other
  millimeter arrays have already yielded masses and surface density
  profiles of many tens of disks by measuring continuum dust emission,
  e.g. \citet{Isella09, Guilloteau11}. ALMA is expected to increase
  the yield to hundreds or thousands of disks on account of its order
  of magnitude better sensitivity and angular resolution \citep{Williams11}.\\
    
\item STEPs may be diverse objects with compositions that reflect
  initial conditions, formation mechanism, and environment.
  Sub-Earths close to their parent stars may be rich in water and
  other volatiles if they originated on wider orbits past the ``snow
  line'' and subsequently migrated or were scattered inwards.
  However, stellar XUV heating and winds are expected to remove their
  atmospheres (Fig. \ref{fig:escape}), and, around solar-type stars,
  the silicate mantles of these planets may evaporate, leaving bare
  iron cores.  Indeed, some sub-Earths may be the product of
  evaporation of more massive planets.  Observations of nearby
  transiting sub-Earths by JWST may be able to discriminate between
  tidally-locked planets lacking atmospheres, which will be hotter and
  brighter, and those with atmospheres, which will be fainter and
  perhaps undetectable
  (Fig. \ref{fig:jwst}).\\
    
\end{itemize}
    
The immediate scientific return from the study of sub-Earths will be
tests of models of planet formation and evolution.  Descriptions
of their occurrence and distributions with mass and orbital period is
essential for a complete description of the planetary kingdom, and any
over-arching theory must explain them. The possible role of
subterrestrial planets as habitats for life should also not be
overlooked.  Although sub-Earths on very close orbits may not be
suitable abodes for life, those further out may orbit in the
circumstellar habitable zone and retain atmospheres and water.  We who
inhabit a comparatively small planet around a ``dwarf"
star should not presume that one Earth mass is the optimum for life.\\
    
\begin{acknowledgements}
  This research has made use of the NASA Exoplanet Archive, which is
  operated by the California Institute of Technology, under contract
  with the National Aeronautics and Space Administration under the
  Exoplanet Exploration Program. Some of the data presented in this
  paper were obtained from the Mikulski Archive for Space Telescopes
  (MAST). STScI is operated by the Association of Universities for
  Research in Astronomy, Inc., under NASA contract NAS5-26555. Support
  for MAST for non-HST data is provided by the NASA Office of Space
  Science via grant NNX09AF08G and by other grants and contracts. This
  paper includes data collected by the Kepler mission. Funding for the
  Kepler mission is provided by the NASA Science Mission directorate.
  EG acknowledges support from NASA grants NNX10AI90G and NNX11AC33G.
  We thank Y. Kokubo for providing Fig. 5, E. Kite for commenting on
  an earlier version of this manuscript and E. Petigura and
  A. Howard for helpful discussions.
\end{acknowledgements}
    
\bibliographystyle{aps-nameyear} 
\bibliography{references_revised_v3}

\end{document}